\documentclass[twocolumn,preprintnumbers,amsmath,amssymb,floatfix,superscriptaddress,nofootinbib]{revtex4-1}

\usepackage{amsmath,amssymb}
\usepackage{graphicx}
\usepackage{dcolumn}
\usepackage{bm}
\usepackage{verbatim}
\usepackage{slashed}
\usepackage{cancel}
\usepackage{float}
\usepackage{enumerate}
\usepackage{setspace}
\usepackage{multirow}

\usepackage{enumitem}
\setlist{nolistsep}

\usepackage{latexsym}
\usepackage{epsfig,amsmath,graphics}
\usepackage{listings}
\usepackage{slashed}
\usepackage{comment,latexsym} 
\usepackage{bm}
\usepackage{tabularx}
\usepackage{dcolumn}
\usepackage{color}

\definecolor{colorLink}{rgb}{0.7,0,0}
\definecolor{colorCite}{rgb}{0,.7,0}
\definecolor{colorURL}{rgb}{0,0,0.7}
\usepackage[colorlinks=true,linktocpage=true,linkcolor=blue,citecolor=colorCite,urlcolor=colorURL]{hyperref}

\usepackage{cleveref}
\crefname{table}{Table}{Tables}
\crefname{equation}{Eq.}{Eqs.}
\crefname{appendix}{App.}{Apps.}
\crefname{section}{Sec.}{Secs.}
\crefname{figure}{Fig.}{Figs.}

\usepackage{tikz}
\usetikzlibrary{arrows,shapes}
\usetikzlibrary{trees}
\usetikzlibrary{matrix,arrows}
\usetikzlibrary{positioning}
\usetikzlibrary{calc,through}
\usetikzlibrary{intersections}
\usetikzlibrary{decorations.pathreplacing}
\usetikzlibrary{decorations.pathmorphing}
\usetikzlibrary{decorations.markings}
\usepackage{pgffor}
\usetikzlibrary{tikzmark,calc}
\tikzset{
    vector/.style={decorate, decoration={snake}, draw},
    provector/.style={decorate, decoration={snake,amplitude=2.5pt}, draw},
    antivector/.style={decorate, decoration={snake,amplitude=-2.5pt}, draw},
    fermion/.style={draw=black, postaction={decorate},decoration={markings,mark=at position .55 with {\arrow[draw=black]{>}}}},
    fermionbar/.style={draw=black, postaction={decorate},decoration={markings,mark=at position .55 with {\arrow[draw=black]{<}}}},
    fermionnoarrow/.style={draw=black},
    doublefermionnoarrow/.style={double,draw=black},
    gluon/.style={decorate, draw=black,decoration={coil,amplitude=4pt, segment length=5pt}},
    scalar/.style={dashed,draw=black, postaction={decorate},decoration={markings,mark=at position .55 with {\arrow[draw=black]{>}}}},
    scalarbar/.style={dashed,draw=black, postaction={decorate},decoration={markings,mark=at position .55 with {\arrow[draw=black]{<}}}},
    scalarnoarrow/.style={dashed,draw=black},
    doublescalarnoarrow/.style={dashed,double,draw=black},
    electron/.style={draw=black, postaction={decorate},decoration={markings,mark=at position .55 with {\arrow[draw=black]{>}}}},
    bigvector/.style={decorate, decoration={snake,amplitude=4pt}, draw},
}

\def\be{\begin{equation}}
\def\ee{\end{equation}}
\def\bea{\begin{eqnarray}}
\def\eea{\end{eqnarray}}

\newcommand{\SM}{\scriptscriptstyle \text{SM}}
\newcommand{\DM}{\scriptscriptstyle \text{DM}}
\newcommand{\mW}{m_{\scriptscriptstyle W}}

\definecolor{colorTC}{rgb}{.2,.7,.2}

\definecolor{colorRTD}{rgb}{.2,.2,.7}

\definecolor{colorML}{rgb}{.2,.7,.7}

\begin{document}

\vspace*{-30mm}

\title{Freezing-in the Hierarchy Problem}
\author{Timothy Cohen}
\affiliation{\,Institute of Theoretical Science, University of Oregon, Eugene, OR 97403 \vspace{2pt}}
\author{Raffaele Tito D'Agnolo}
\affiliation{\,Theory Group, SLAC National Accelerator Laboratory, Menlo Park, CA 94025\vspace{2pt}}
\author{Matthew Low\,}
\affiliation{\,School of Natural Sciences, Institute for Advanced Study, Princeton, NJ 08540 \vspace{3pt}}

\begin{abstract}
\begin{centering}
{\bf Abstract}\\[4pt]
\end{centering}
\noindent Models with a tiny coupling $\lambda$ between the dark matter and the Standard Model, $\lambda \sim v/M_\text{Pl}\sim 10^{-16}$, can yield the measured relic abundance through the thermal process known as freeze-in. We propose to interpret this small number in the context of perturbative large $N$ theories, where couplings are suppressed by inverse powers of $N$. Then $N \sim M_{\rm Pl}^2/v^2$ gives the observed relic density.  Additionally, the ultimate cutoff of the Standard Model is reduced to $\sim 4\,\pi\, M_\text{Pl}/\sqrt{N} \sim 4\, \pi\, v$, thereby solving the electroweak hierarchy problem. These theories predict a direct relation between the Standard Model cutoff and the dark matter mass, linking the spectacular collider phenomenology associated with the low gravitational scale to the cosmological signatures of the dark sector. The dark matter mass can lie in the range from hundreds of keV to hundreds of GeV. Possible cosmological signals include washing out power for small scale structure, indirect detection signals from dark matter decays, and a continuous injection of electromagnetic and hadronic energy throughout the history of the Universe.
\end{abstract}

\vspace*{1cm}

\maketitle

\begin{spacing}{1.15}
Dark matter (DM) accounts for $80\%$ of the matter in our Universe, but its microscopic origin is still unknown. If the DM is coupled to the Standard Model (SM) thermal bath, a variety of mechanisms to produce the observed relic abundance are possible. The existence of a connection between the DM and the SM implies that measurements of these couplings today can teach us about the dynamics of the early Universe.  One compelling thermal scenario is ``freeze-in''~\cite{Hall:2009bx}, see~\cite{Bernal:2017kxu} for a recent review.   Initially, the SM is reheated and thermalizes, while the DM sector is not  and therefore has a negligible energy density.  As the Universe expands, the very small couplings to the SM mediate out-of-equilibrium processes that generate a DM number density. The DM production ceases once the SM bath temperature $T$ becomes of order of the mass of the lightest SM particle that interacts with the DM, thereby freezing-in the relic density.

A key feature of freeze-in models is their very small coupling between the SM and DM.  In this paper, we interpret this aspect in a new context, which leads us to find an interesting connection with the electroweak hierarchy problem. If the number $N$ of new states in the dark sector is large, but the theory is under perturbative control, the DM couplings to the SM must scale with an appropriate inverse power of $N$.  Here we posit that this is the origin of the tiny couplings required for a viable model of freeze-in.  We will show that $N\sim M_{\rm Pl}^2/v^2$ reproduces the observed relic density, where $v$ is the scale of electroweak symmetry breaking.

Theories with a huge number of new states are exciting for another reason. Gravity interacts with all the new degrees of freedom.  Therefore, graviton-graviton scattering at energy $E$ receives quantum corrections proportional to powers of $N\, E^2/\big(16\, \pi^2\, M_\text{Pl}^2\big)$~\cite{Dunbar:1994bn, Calmet:2007je, Anber:2011ut}.  We expect the ultimate scale where gravity is modified to be reduced to~\cite{Dvali:2007hz, Dvali:2007wp, Dvali:2007iv, Dvali:2009ne, Dvali:2009fw}
\vspace{-6pt}
\begin{align}
\Lambda_\text{UV} \sim 4\,\pi\,\frac{M_\text{Pl}}{\sqrt{N}}\, .
\label{eq:LambdaUV}
\end{align}
If $N\sim M_{\rm Pl}^2/v^2$, this is a dramatic reduction in the size of the ultimate cutoff of the SM, and having $N$ dark sector states solves the hierarchy problem.\footnotemark\setcounter{footnote}{0} In this paper we show that viable DM models exist with $N \sim 10^{15-35}$, implying  $\Lambda_\text{UV} \sim 1-10^{10}\text{ TeV}$.  

If $\Lambda_\text{UV}$ is within reach of the LHC or a future machine, then spectacular collider signatures associated with the gravitational sector would likely be discovered~\cite{Dvali:2009ne}. On the other hand, it is unsurprising that having $\Lambda_\text{UV} \sim v$ comes at a price.  In particular, new physics associated with gravity should appear at $\Lambda_\text{UV}$, which can break baryon and/or lepton number, and can induce flavor-changing and/or CP-violating processes. One way to avoid the suite of associated constraints is to impose additional structure in the UV theory above $\Lambda_\text{UV}$.  Alternatively, it is perfectly viable to lift $\Lambda_\text{UV}$ above  $v$, but at the expense of tuning the weak scale.  For a given model, the freeze-in relic density calculation maps $N$ (and therefore $\Lambda_\text{UV}$) onto a choice for the DM mass, as discussed in \cref{sec:relic} and \cref{sec:results}. This relation is one of the most appealing features of our framework since it implies that this class of theories is quite predictive: once we fix the large $N$ scaling of the couplings, the DM mass determines both the strength of its interactions with the SM and $\Lambda_\text{UV}$. 

Since at first blush introducing $N \sim 10^{\text{many}}$ new degrees of freedom might seem extreme, it is worth briefly commenting on the connection to extra dimensional scenarios.  One phenomenologically relevant example comes from introducing a mm-sized compactified extra dimension~\cite{ArkaniHamed:1998rs, Antoniadis:1998ig, ArkaniHamed:1998nn, Giudice:1998ck}, which also modifies gravity at the TeV scale. In the 4D effective theory, this can be viewed as a consequence of a large number of Kaluza-Klein states, comparable to the number of species in our dark sector. In this sense, the equivalence of the large $N$ solution to the hierarchy problem can be made sharp.  A similar emergence of large $N$ also occurs in calculable models of the AdS/CFT correspondence~\cite{Maldacena:1997re, Witten:1998qj, Gubser:1998bc, Aharony:1999ti}.  We emphasize that here we do not attempt to explain the origin of the $N$ dark sector states using an explicit extra dimensional construction, but making this connection precise, and understanding what theoretical restrictions it imposes, would be an interesting task for future work.

The rest of the paper is organized as follows. In \cref{sec:scalings}, we write down two example models and discuss their large $N$ scalings.  Then in \cref{sec:relic}, we present the calculation of the freeze-in abundance, and emphasize the modifications due to having $N$ species.  Our numerical results and a discussion of the related phenomenology are given in \cref{sec:results}, followed by a brief outlook.

\footnotetext{Introducing a large number of new degrees of freedom is one of the simplest known solution to the hierarchy problem~\cite{Dvali:2007hz, Dvali:2007wp, Dvali:2007iv, Dvali:2009ne, Dvali:2009fw}.  The novelty here is the connection to freeze-in.}

\setcounter{section}{1}
\section{Models}
\label{sec:scalings}
In this Section, we introduce two example models that realize the goals set forth above. We also discuss the requirements imposed by perturbativity on their couplings. For context, in~\cite{tHooft:1973alw}, 't~Hooft demonstrated that in an U$(N)$ gauge theory with gauge coupling $g$, holding the combination $g^2\, N$ fixed while taking the limit $N \to \infty$ maintains the validity of the perturbative expansion.  This leads to a natural division into planar diagrams, which scale with powers of $g^2\,N$, and non-planar diagrams which are suppressed by powers of $1/N$ with respect to the planar graphs.  The perturbative expansion of the theory is reorganized in terms of the small parameter $1/N$~\cite{tHooft:1973alw}.

We note that the relation between the topology of diagrams and the $1/N$ expansion is a feature that we do not find in our simple models. This is due to the absence of fields that are the analog of the gluons in~\cite{tHooft:1973alw}.  For both models presented here, we derive the minimal $N$ scaling of the coupling needed to maintain perturbativity.  While what follows maintains the same spirit, our arguments differ from 't~Hooft's approach in the details.

\vspace{30pt}
\noindent {\bf Scalar Model:} Our first example has $N$ real scalar DM candidates $\phi_\alpha$ coupled to the SM via the Higgs portal
\be
\mathcal{L} \supset - \lambda_\phi\, |H|^2\, \sum_{\alpha=1}^N \phi_\alpha^2\, ,
\label{eq:ScalarModel}
\ee
whose stability is maintained by enforcing a $\mathbb{Z}_2$ symmetry: $\phi_\alpha \rightarrow - \phi_\alpha$. In the parameter space of interest, $m_{\phi_\alpha}<m_h/2$, and the DM abundance is set by the decays $h\rightarrow \phi_\alpha\,\phi_\alpha$ after the electroweak phase transition. 
As discussed in \cref{sec:results}, the $\phi_\alpha$ do not introduce $N$ hierarchy problems, given the tiny size of the coupling $\lambda_\phi$. 

Every diagram constructed from the interaction in \cref{eq:ScalarModel} contains a factor $r$, which scales as 
\be
r\sim \lambda_\phi^V N^{L_\phi},
\label{eq:ScalarCounting1}
\ee
where $V$ is the total number of vertices, and $L_\phi$ is the number of closed $\phi_\alpha$ loops. To make the structure of the perturbative expansion more manifest we would like to express $L_\phi$ in terms of $V$.   

For any diagram, if we remove all external Higgs lines and internal Higgs propagators, we are left with $n$ disconnected subdiagrams (where $n \geq 1$) consisting only of $\phi_\alpha$ lines.  We can then write 
\be
L_\phi = \sum_{i=1}^n L_{\phi,i} = \sum_{i=1}^n (V_i - P_i + 1)
= V - P + n,
\ee
where $V_i$ and $P_i$ are the number of vertices and edges respectively in the $i^\text{th}$ subdiagram, $V\equiv\sum_{i=1}^n V_i$ and $P\equiv\sum_{i=1}^n P_i$. Then \cref{eq:ScalarCounting1} becomes
\be
r \sim \big(\lambda_\phi\, N\big)^V N^{- P+n}\,.
\label{eq:ScalarCounting2}
\ee
Since every subdiagram has at least one edge, $P\geq n$ and perturbativity is maintained by imposing $\lambda_\phi \lesssim 1/N$.  In particular, the leading diagrams scale as $(\lambda_\phi\, N)^V$ and consist of a single vertex per $\phi_\alpha$ loop\footnote{In this theory, the leading diagrams renormalize the Higgs mass and the cosmological constant.  If one is willing to fine-tune the Higgs mass, then perturbativity  can be maintained with a weaker scaling $\lambda_\phi \sim 1/\sqrt{N}$.}
\be
\begin{tikzpicture}[line width=1.0 pt, scale=1.0,baseline={([yshift=-1ex]current bounding box.center)}]
  \draw[fill=black]          (0,0) circle (0.05);
  \draw[doublescalarnoarrow] (0,0) arc ( 90:450:-0.6);
  \draw[scalarnoarrow]       (0,0) -- ( 1,-0.5);
  \draw[scalarnoarrow]       (0,0) -- (-1,-0.5);
  \node at (   0, 0.9) {$\phi_\alpha$};
  \node at (-1.3,-0.7) {$h$};
  \node at ( 1.3,-0.7) {$h$};
  \node at (   0,-0.4) {$\lambda_\phi$};
\end{tikzpicture} \hspace{5pt} \sim\hspace{3pt} \lambda_\phi\, N\,.
\ee

\vspace{30pt}
\noindent {\bf Fermion Model:} Our second example introduces a fermionic DM candidate $\psi_\alpha$, that couples to the SM through the neutrino portal
\be
\mathcal{L} \supset -\, y_\psi\, L_e\, H\, \sum_{\alpha=1}^N \psi_\alpha\, ,
\label{eq:FermionModel}
\ee
where $L_e$ is the SM lepton doublet containing the electron. We have chosen to couple only to the electron doublet mainly for simplicity. Given the relevant size of $y_\psi$ discussed below, generalizing this coupling would have a minimal impact on the DM phenomenology, and as such we leave exploring this larger parameter space to future work. The small size of $y_\psi$ also insures that the $\psi_\alpha$ are stable on cosmological timescales, as discussed in more detail in \cref{sec:results}.

In this model,\footnote{In this model, there is also a potential issue with maintaining unitarity at tree-level, for example in the process $h\,\nu \rightarrow h\,\nu$.  The same 't~Hooft scaling we derive at loop-level, also maintains the finiteness of the tree processes in the large $N$ limit.} each diagram contains a factor
\be
r\sim y_\psi^V\, N^{I_\psi},
\label{eq:FermionCounting1}
\ee
where $V$ is the total number of vertices, as above, and $I_\psi$ is the number of internal $\psi_\alpha$ propagators.  As there is only one $\psi_\alpha$ line coming out of each vertex, we can write
\be
V = 2\, I_\psi + E_\psi,
\ee
where $E_\psi$ are external $\psi_\alpha$ lines.  Then 
\be
r\sim\big(y_\psi^2\, N\big)^{V/2} N^{-E_\psi/2}.
\label{eq:FermionCounting2}
\ee
The 't~Hooft coupling is $y_\psi^2\, N$ and perturbativity is maintained by requiring $y_\psi \lesssim 1/\sqrt{N}$.  The leading diagrams scale as $y_\psi^2\, N$ and have no external $\psi_\alpha$ lines
\be
\begin{tikzpicture}[line width=1.0 pt, scale=1.0,baseline={([yshift=-1ex]current bounding box.center)}]
  \draw[doublefermionnoarrow] (0,0) arc (  0:180: 0.6);
  \draw[fermionnoarrow]       (0,0) arc (  0:-180: 0.6);
  \draw[scalarnoarrow]        (0,0) -- (1.2,0);
  \draw[scalarnoarrow]        (-1.2,0) -- (-2.4,0);
  \draw[fill=black] (   0,0) circle (0.05);
  \draw[fill=black] (-1.2,0) circle (0.05);
  \node at ( 1.4,0) {$h$};
  \node at (-2.6,0) {$h$};
  \node at (-0.6,-0.8) {$\nu$};
  \node at (-0.6, 0.9) {$\psi_\alpha$};
  \node at ( 0.3, 0.3) {$y_\psi$};
  \node at (-1.5, 0.3) {$y_\psi$};
\end{tikzpicture} \hspace{5pt} \sim\hspace{3pt} y_\psi^2\, N\,.
\ee
As we will show in the next two sections, both the scalar and fermion model can realize the measured relic abundance via freeze-in, while satisfying all the relevant phenomenological constraints.
\section{Relic Density}
\label{sec:relic}
Now that we have discussed the interplay between the requirement of perturbativity and the large $N$ scalings of the couplings, we turn to the calculation of the thermal freeze-in relic abundance.  The goal of this section is to review the mechanism~\cite{Hall:2009bx} with an emphasis on dark sectors with a large number of new species.  We begin with a discussion of the parametrics, which are then derived using the full Boltzmann equations.

The freeze-in abundance of a single species due to the decay of an SM particle with mass $m_{\SM}$ is
\be
Y_{N = 1} \equiv \frac{n_{N = 1}}{s} \sim \lambda^2 \frac{M_{\rm Pl}}{m_{\SM}}\, ,
\ee
where $\lambda$ is the coupling between the SM and the new particle. The observed relic abundance is reproduced when $Y\sim v^2/(M_{\rm Pl}\, m_{\DM})$, implying that\footnote{A more accurate way to estimate the measured DM yield is $Y_{\rm obs}\sim T_{\scriptsize\text{EQ}}/m_{\DM}$, where $T_{\scriptsize\text{EQ}}\sim {\rm eV}$ is the SM temperature at matter-radiation equality. The approximate accident $T_{\scriptsize\text{EQ}} \sim v^2/M_{\rm Pl}$ is one common way of expressing the WIMP miracle~\cite{Lee:1977ua, Kolb:1990vq, Gondolo:1990dk, Jungman:1995df}, and we exploited it here as well since it yields simple scaling relations.  If one wants to make quantitative comparisons, we recommend using \cref{eq:approxrelic}.} 
\be
\lambda \sim \frac{v}{M_{\rm Pl}}\sqrt{\frac{m_{\SM}}{m_{\DM}}}\, . 
\label{eq:LambdaFI}
\ee
In our scalar scenario defined in \cref{eq:ScalarModel}, we have shown that the coupling should scale at most as $1/N$.  Even when this scaling is fixed, there is a range of possibilities for the many particle abundance $Y_N$. If the couplings are diagonal, then only the channels $\text{SM} \rightarrow \text{DM}_\alpha\,\text{DM}_\alpha$ are active so that the sum over the particles in the final state gives $Y_N= N\, Y_{N = 1} $, and \cref{eq:LambdaFI} becomes
\be
N\sim \frac{M_{\rm Pl}^2}{v^2}\frac{m_{\DM}}{m_{\SM}}\, \qquad \text{(scalar model)} . 
\label{eq:N}
\ee
To simplify the parametrics, we have assumed that all the dark sector particles have the same mass and coupling to the SM. 

On the other hand, if the couplings are all active, then $\text{SM} \rightarrow \text{DM}_\alpha\,\text{DM}_\beta$ with $\alpha\neq \beta$ are allowed, giving $N(N+1)/2$ decay channels. Note however that this possibility makes sense only in sectors with different masses for the scalars or additional interactions. Putting it all together, again in the simple limit where all the couplings and masses are the same, we have 
\begin{align}
Y_{N=1} &\leq Y_N \lesssim N^2\, Y_{N=1}\notag\\[5pt]
\Longrightarrow \qquad \frac{1}{N}\frac{v}{M_{\rm Pl}}\sqrt{\frac{m_{\SM}}{m_{\DM}}} &\lesssim  \lambda_\phi \lesssim \frac{v}{M_{\rm Pl}}\sqrt{\frac{m_{\SM}}{m_{\DM}}}\, .
\end{align}
We see that in theories where the off-diagonal decay channels are active, we have to suppress the coupling beyond the 't~Hooft scaling in order to reproduce the observed relic density.  

In our fermionic scenario defined in \cref{eq:FermionModel}, DM candidates couple to the SM via a seesaw type mixing with neutrinos. In this case their decay width is suppressed by an extra factor $v^2/m^2_{\DM}$ and a simple possibility is to have couplings $\sim 1/N$ and $N$ decay channels.\footnote{This choice anticipates the phenomenological constraints discussed in the next section when $y_\psi \sim 1/\sqrt{N}$ and there are $N$ active decay channels.} This gives
\be
N\sim \frac{M_{\rm Pl}^2}{m_{\SM}\, m_{\DM}}\, \qquad \text{(fermion model)}\,.
\label{eq:Nmixing}
\ee
We will see in \cref{sec:results} that even though they mix with the neutrinos, these DM candidates can easily be stable on cosmological timescales.

Now we will show how these parametric relations all follow from the Boltzmann equations.  We can start with the equation for a single species with phase space density $f_\alpha$, and make all the usual approximations: we neglect Pauli blocking, $1\pm f_{\SM} \simeq 1$, assume that our new sector starts out without any energy density, $f_{\alpha}\simeq 0$, and take the SM progenitor in thermal equilibrium with the SM bath, $f_{\SM} \simeq e^{-E_{\SM}/T}$. This yields
\be
\frac{\text{d} n_\alpha}{\text{d}t}+3\, H\, n_\alpha \simeq \frac{g_{\SM}\, m_{\SM}^2}{2\,\pi^2}\, T\, K_1\bigg(\frac{m_{\SM}}{T}\bigg) \sum_{\beta=1}^N \Gamma_{\alpha\beta}\, , \label{eq:BEi}
\ee
where $K_1$ is the first modified Bessel function of the second kind, $\Gamma_{\alpha\beta}=\Gamma(\text{SM} \rightarrow \text{DM}_\alpha\,\text{DM}_\beta)$, and $g_{\SM}$ is the number of internal degrees of freedom of the SM particle that is decaying.  When $T> m_{\SM}$ the term on the right hand side leads to an increase in the DM number density, even if it is out of equilibrium. As the temperature lowers to $T\simeq m_{\SM}$, the SM particle number density begins to be exponentially suppressed, effectively stopping DM production. This suggests that  freeze-in is dominated by $T_{\text{FI}}\simeq m_{\SM}$.  However, to estimate the relic density we cannot equate the expansion of the Universe with the rate of the decays at $T_{\text{FI}}$ -- as we would do for freeze-out calculations -- since the decays never enter equilibrium in this scenario. Instead we need to integrate \cref{eq:BEi}, following~\cite{Hall:2009bx}.

If we take all the particles in the hidden sector to have the same mass (or assume that they all eventually decay to the lightest member of their sector), the relic density today is proportional to the sum of their number densities
\be
\Omega_{\DM}\, h^2 \propto \rho_{\DM} = m_{\DM} \sum_{\alpha=1}^N n_\alpha\equiv m_{\DM}\, n\, ,
\ee 
implying that we can compute $\Omega_{\DM}$ by summing over $\alpha$ in \cref{eq:BEi}. 
Integrating the equation between $T=0$ and $T=\infty$, we find 
\be
\Omega\, h^2 \simeq \frac{1.1\times 10^{27}\, g_{\SM}}{g_{*}^{3/2}(m_{\SM})}\frac{m_{\DM}}{m_{\SM}^2}\sum_{\alpha=1}^N\sum_{\beta=1}^N \Gamma_{\alpha\beta}\, ,
\label{eq:approxrelic}
\ee
in good agreement with a full numerical calculation. Here $g_*\simeq g_{*S}$ counts the effective number of degrees of freedom in the SM thermal bath. This derivation shows that our multi-particle yield, $Y_N$, is proportional to the sum of $\Gamma_{\alpha\beta}$ over all the active channels. This is sufficient to derive all the parametric relations discussed at the beginning of this section. 

\section{Phenomenology}
\label{sec:results}
In this section, we present the experimentally viable parameter space for our scalar and fermion models along with the associated signatures. We also make quantitative the relation between the DM mass and the gravitational cutoff. First, we assume that all particles have identical masses, which allows us to clearly illustrate the relevant constraints and the correlation between the DM mass and $\Lambda_\text{UV}$.  We then discuss how the phenomenology is modified in the case that the dark sector masses follow a non-trivial distribution.

\subsection{Equal Masses}
\begin{figure*}[t!]
\begin{center}
\includegraphics[width=0.49\textwidth]{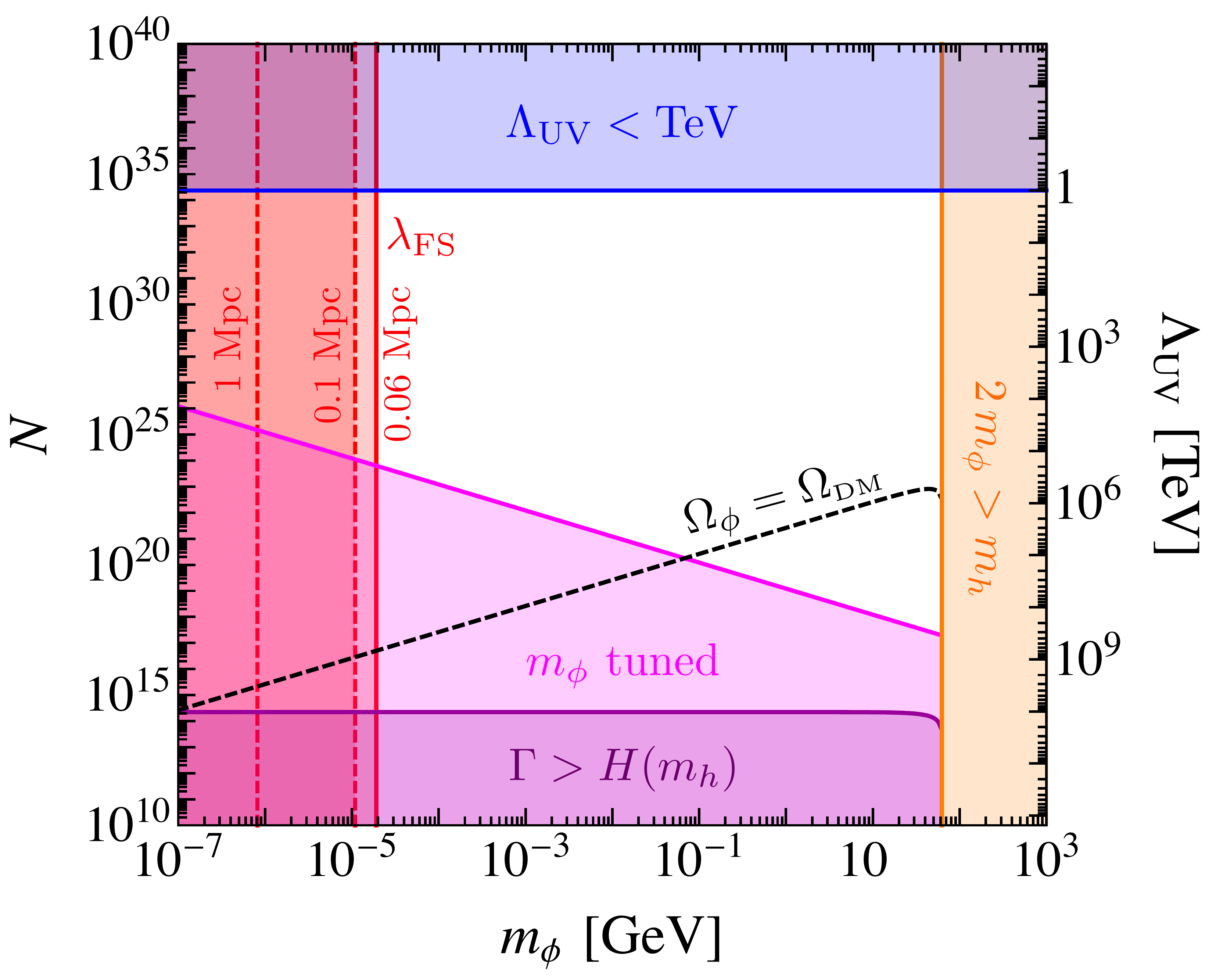} 
\hspace{4pt}
\includegraphics[width=0.49\textwidth]{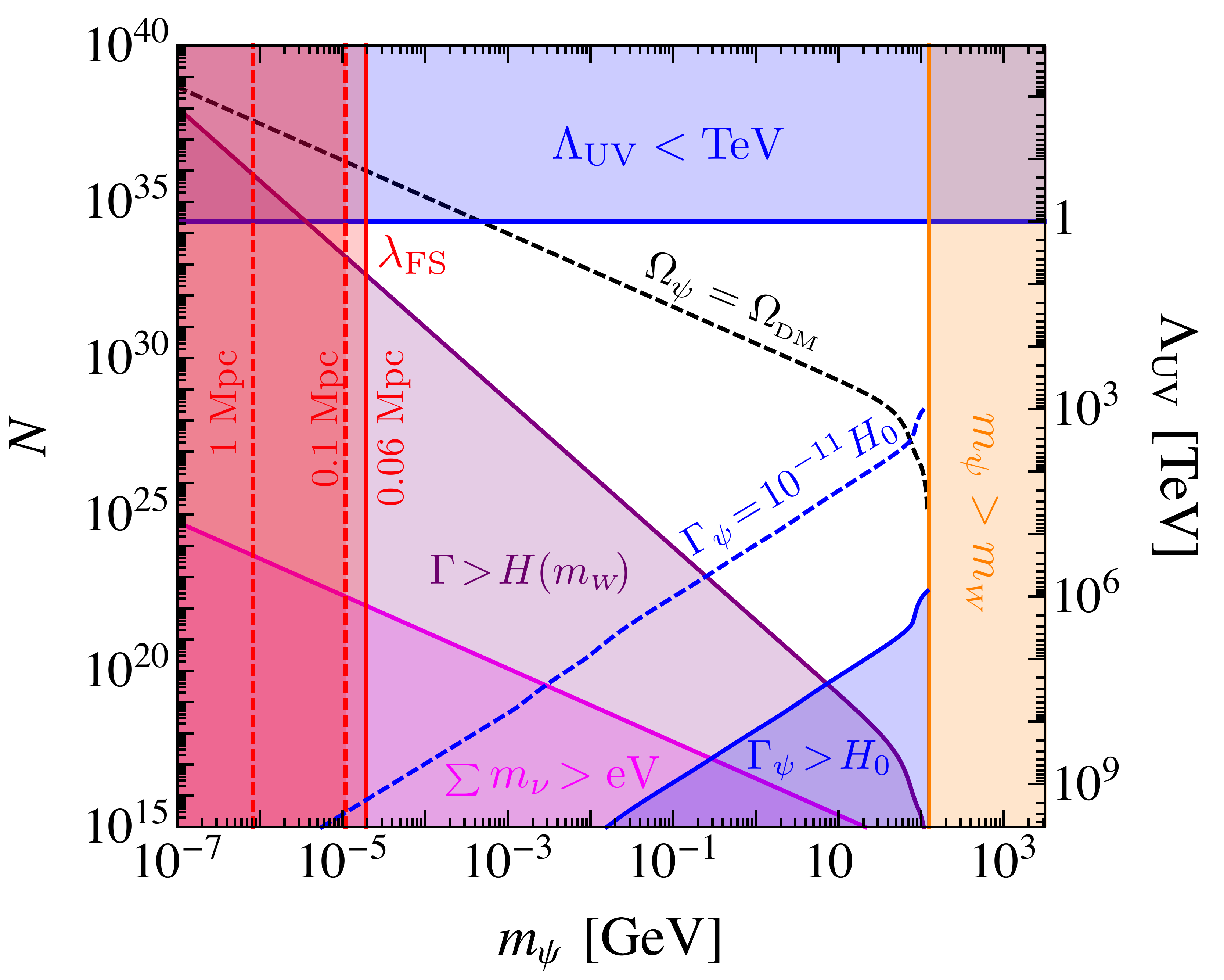} 
\caption{{\it Left}: Phenomenology of the scalar DM model defined by \cref{eq:ScalarModel}. The DM is produced by $h\to \phi_\alpha\,\phi_\alpha$ decays. {\it Right}: Phenomenology of the fermion DM model defined by \cref{eq:FermionModel}. The DM is dominantly produced by $W^\pm\to e^\pm\,\psi_\alpha$ decays.  Note that in the parameter space below the relic density line $\Omega = \Omega_{\DM}$ the DM overcloses the Universe if a standard cosmological history is assumed. The figure is discussed in detail in Section~\ref{sec:results}.} 
\label{fig:twobody}
\end{center}
\end{figure*}

In our first example model, the DM is composed of $N$ scalar particles with masses $m_\phi \lesssim m_h/2$ that couple to the SM through the Higgs portal as in \cref{eq:ScalarModel}. For simplicity, we enforce that the DM is stable, which implies that none of the $\phi_\alpha$ states obtain a vacuum expectation value.  Then the DM relic density freezes-in after the electroweak phase transition from two-body decays of the Higgs, $h\to \phi_\alpha\, \phi_\alpha$. The abundance produced during the electroweak symmetric phase from the process $H\,H^\dag \to \phi_\alpha\, \phi_\alpha$ is negligible.  If only diagonal decays are active, then saturating the 't~Hooft scaling of the coupling, $\lambda_\phi=1/N$, yields the observed relic density along the black curve in the left panel of \cref{fig:twobody}.  Furthermore, this figure shows that the parameter space for the scalar model has a lower bound on the cutoff, $\Lambda_\text{UV} \gtrsim 10^6$~TeV. In this model, $\Lambda_\text{UV}\sim m_\phi^{-1/2}$ as determined by \cref{eq:N}, such that the cutoff is lowest ($N$ largest) for the largest kinematically allowed DM mass, in this case set by $m_h$.

In our second example model, the DM is made up of $N$ fermions with masses $m_\psi \lesssim \mW$.  They couple to the SM via the neutrino portal in \cref{eq:FermionModel}, \emph{i.e.}, the same couplings as right-handed neutrinos. As discussed above, perturbativity requires $y_\psi \lesssim 1/\sqrt{N}$ and the process $W^\pm\rightarrow \psi_i\,e^\pm$ dominantly determines the freeze-in abundance. Saturating the perturbativity bound on $y_\psi$ can never yield the measured relic density without violating phenomenological constraints. This is straightforward to see by taking~\cref{eq:approxrelic} and evaluating it at $T_{\text{FI}}\simeq \mW$, which yields $
\Omega_\psi\, h^2 \simeq 3\times 10^{24}\, m_\psi/\mW$, implying $m_\psi \simeq 10^{-15}$~eV.  As discussed below, such a light mass is ruled out by measurements of structure formation.  

To find phenomenologically viable parameter space, we can either reduce the number of decay channels, or note that if $y_\psi \ll N^{-0.5}$ it is easy to reproduce the correct relic density -- for concreteness, results are shown in \cref{fig:twobody} for the (arbitrary) choice $y_\psi= N^{-0.9}$.  We see that the cutoff can easily be brought down to the weak scale or even below with the fermion model. This is shown in the right panel of \cref{fig:twobody}. In this case, the mixing with neutrinos impacts the relation between $N$ and the DM mass (see \cref{eq:Nmixing}), implying a scaling $\Lambda_\text{UV}\sim m_\psi^{1/2}$.  

Next we turn to the experimental bounds of our parameter space, most of which apply to both models.  As $N$ becomes large enough, the gravitational cutoff passes below a TeV, which is in conflict with multiple observations. At small $N$, the coupling between the dark sector and the SM becomes too large, and we are no longer in the freeze-in regime. In the plots, this is denoted by the shaded region where $\Gamma > H(m_{\SM})$, \emph{i.e.}, the decay rate that populates the dark sector enters equilibrium at the freeze-in temperature. In the fermion model, another slice of small $N$ parameter space is excluded if $\Gamma_\psi > H_0$, the requirement of stability on timescales comparable to the age of the Universe. 

The DM mass is also bounded from above and from below in both models. As $m_{\DM}$ approaches the weak scale, the SM to dark sector decays become kinematically forbidden.\footnote{We can still populate the dark sector through inverse decays, but we do not consider this regime here.} For simplicity, we derive this kinematic bound in the zero temperature limit, since at $T_{\text{FI}}\simeq m_{\SM}$ thermal corrections are negligible for our purposes. 

In both models, one lower bound on the mass follows from considering structure formation.  In order to derive a constraint, we need to estimate the free-streaming length.  This is done assuming that all of DM is produced at $T_\text{FI}$ with an energy determined by the kinematics of the decays.  We do not include any self-interactions in the dark sector. This set of assumptions implies that initially the DM free-streams relativistically, starting from the time of production $t_p$ ($T\sim v$). After $t_\text{NR}$ (when $T\sim m_{\DM}$) it continues to free-stream non-relativistically until matter-radiation equality at $t_\text{EQ}$.  This yields a free-streaming length
\begin{align}
\hspace{-6.5pt}\lambda_{\text{FS}} &= a_0\, c\int_{t_p}^{t_{\text{NR}}}\!\!\text{d}t \,\frac{1}{a(t)} + a_0 \int_{t_\text{NR}}^{t_\text{EQ}}\!\!\text{d}t\,\frac{v_{\DM}(t)}{a(t)} \nonumber \\[5pt]
&\simeq 0.1 \text{ Mpc} \; \sqrt{\frac{g_*\big(\text{keV}\big)}{g_*\big(m_{\DM}\big)}}\frac{\rm keV}{m_{\DM}}\bigg[1+\log\left(\frac{m_{\DM}}{\rm eV}\right)\bigg]\,,\!\!
\end{align}
where $a(t)$ is the scale factor of the Universe as a function of time (we assume a standard cosmological history), $a_0$ is the scale factor today, $c$ is the speed of light, and $v_{\DM}(t)$ is the red-shifting DM velocity.  Given that after matter-radiation equality the free-streaming length grows only logarithmically, we can safely neglect this late-time contribution.    In \cref{fig:twobody}, we show the regions where $\lambda_{\text{FS}} \gtrsim  0.06 \text{ Mpc}, 0.1 \text{ Mpc}, \text{ and } 1 \text{ Mpc}$.  Currently, $m_{\DM}\lesssim 0.01$~MeV is in tension with observations from Lyman-$\alpha$ forest data~\cite{Irsic:2017ixq, Viel:2013apy}.  

It might seem that the tiny coupling between the dark sector and the SM precludes any hope of a positive signal other than the possible suppression of the small scale power spectrum. However, since all of DM is unstable in the fermion model there are strong constraints from indirect detection.  For masses above $10$~GeV, the bounds from a dedicated analysis of Fermi gamma rays~\cite{Cohen:2016uyg} are relevant, while below a combination of gamma-ray and $X$-ray data are more important~\cite{Essig:2013goa}.  While a re-interpretation for the decays $\psi_\alpha \rightarrow W^{(*)}\, \ell, Z^{(*)}\, \nu, h^{(*)}\, \nu$ is left for future work, we show the line through our parameter space in the right panel of \cref{fig:twobody} where $\Gamma_{\psi} \simeq 10^{-28}/\text{s}$. Above this line DM is more long-lived and the parameter space is unconstrained, while a rough reinterpretation of~\cite{Cohen:2016uyg} suggests that in our model everything below this line, $m_\psi \gtrsim 70$~GeV, is excluded. That these models could be discovered using indirect detection is remarkable in view of the smallness of the coupling between DM and the SM $1/N \sim v^2/M_{\rm Pl}^2 \sim 10^{-33}$, and leaves open the possibility of exploring more of our parameter space with future indirect detection experiments, see \emph{e.g.}~\cite{Gaskins:2016cha} for a discussion.

This completes the picture of the possible signals and the bounds from experiment.  However, there is one important theoretical consideration to discuss for the scalar model.  When the coupling $\lambda_\phi$ becomes sufficiently large, the quartic $\phi$-$H$ coupling in \cref{eq:ScalarModel} introduces a quadratic correction to the $\phi_\alpha$ mass at one-loop. The area shaded in magenta in the left panel of \cref{fig:twobody} indicates a naive tuning bound where $\lambda_\phi\, \Lambda_\text{UV}^2/(16\, \pi^2) \gtrsim m_\phi^2$.  In particular, $m_\phi \simeq 0.1$~GeV is especially interesting since here we simultaneously reproduce the relic density, and generate the correct $\phi_\alpha$ mass at one-loop. In the 't~Hooft limit, $\lambda_\phi= 1/N$, the tuning of the Higgs mass is not considerably worsened by the presence of $N$ $\phi$'s, since they effectively behave as a single particle with an $\mathcal{O}(1)$ coupling, giving a contribution to the Higgs mass $\delta m_h^2 \sim \Lambda_\text{UV}^2/(16\, \pi^2)$.   In the case of fermions, since we are going beyond the 't Hooft limit, there is no appreciable tuning introduced by \cref{eq:FermionModel}. From the point of view of the SM, our $N$ fermions behave as a single particle with an $\mathcal{O}(1/N^{0.4})$ coupling.   Also, their tree-level contribution to the sum of the SM neutrino masses is well below an eV in most of the parameter space, as shown in the right panel of \cref{fig:twobody}.

\subsection{Non-trivial Mass Distributions}
A non-degenerate dark sector mass distribution has the immediate consequence that $N-1$ species are unstable, which could yield new signals in indirect detection, CMB experiments, and measurements of light element abundances, implying very tight constraints.  This shares many of the phenomenological consequences of dynamical DM~\cite{Dienes:2011ja, Dienes:2011sa, Dienes:2016kgc}.

The experimentally relevant quantity is the differential distribution of the energy density as a function of the lifetime
\be
\frac{\text{d}\rho_{\DM}}{\text{d}\tau}=\frac{\text{d}\rho_{\DM}}{\text{d} m}\left|\frac{\text{d}m}{\text{d}\tau}\right| \, .
\ee
Using this expression requires inverting the relation between the decay width and the mass to obtain $m(\tau)$. In general this is challenging to do analytically since the decay width is a convolution over the distribution of masses.  To gain some insight, we assume the case of a very narrow uniform distribution.  If the mass distribution is broad, the model is badly excluded by constraints on energy injection in the early history of the Universe. If the distribution is narrow, we can expand the decay width as a function of the dark sector mass splittings 
\be
\!\!\Gamma\big(\text{DM}_\alpha \to \text{DM}_\beta+X\big) = \Gamma^{(n)} \times\left(\frac{m_\alpha - m_\beta}{m_{\SM}}\right)^n+ \dots\,, \label{eq:width}
\ee
where $n$ is the leading non-zero power in terms of the small mass splitting $(m_\alpha - m_\beta)/m_{\SM}$ and depends on the model: for the scalar model, $X=h^{(*)}$, $m_{\SM}=m_h$, $n=3$, and $\Gamma^{(3)} \sim y_{\SM}^2/(128\, \pi^3\, N^2)(v^2/m_h)$, where $y_{\SM}$ can be an SM Yukawa coupling or a gauge loop factor.  The total width for $\text{DM}_\alpha$ can be easily obtained by integrating \cref{eq:width} over the uniform distribution of masses. The last step needed to obtain $\text{d}\rho_{\DM}/\text{d}\tau$ requires noticing that since all the DM particles have nearly the same mass, they are also produced nearly at the same rate from SM decays so that we can assume that their number densities are all equal. As a consequence, the heaviest $\text{DM}_\alpha$ has the largest initial energy density $\text{d}\rho_{\DM}/\text{d}m\sim m$. This together with \cref{eq:width} implies close to stability
\be
\frac{1}{\rho_{\DM}}\frac{\text{d}\rho_{\DM}}{\text{d}\tau} \sim \tau^{-\frac{n+2}{n+1}}\, ,
\ee
peaked on the shortest allowed lifetime. 
Clearly this result strongly depends on the choice of mass distribution. For example, choosing a Normal distribution or a Wigner semicircle distribution implies that $\text{d}\rho_{\DM}/\text{d}\tau$ would be peaked around a typical lifetime allowing for more interesting phenomenological possibilities. For instance a Gaussian tail of the energy density could give continuous energy injection, \emph{e.g.} between BBN and the present epoch.  While a detailed study is beyond the scope of this work, this is one of the more striking phenomenological signatures that can be motivated by models that freeze-in the hierarchy problem.
\newpage
\section{Outlook}
\label{sec:outlook}
In this paper, we have made the connection between the large $N$ solution to the electroweak hierarchy problem and the freeze-in mechanism to generate the relic abundance of DM.  We provided two simple models both of which introduce $N$ dark sector states, and an interaction with the SM whose size is determined by a 't~Hooft coupling.  We showed that it is indeed possible to achieve a phenomenologically viable scenario with interesting signatures. For example, a continuous injection of energy throughout the history of the Universe could be realized.

In the most interesting parameter space where the full hierarchy problem is solved, the ultraviolet cut-off is within reach of current and/or future colliders.  Therefore, it should be possible to produce the $N$ states constituting dark matter through their gravitational interactions. Furthermore, in the case of the fermionic model, we can also directly access the freeze-in  coupling of dark matter through indirect detection, by measuring its lifetime. This is a rather unique example in which dark matter interacts only gravitationally or through extremely feeble couplings, but we can still produce it directly and probe its freeze-in interactions. 

The literature does contain ideas with a similar spirit, \emph{e.g.} a class of large $N$ DM models~\cite{Dvali:2009fw}, dynamical DM~\cite{Dienes:2011ja, Dienes:2011sa, Dienes:2016kgc}, and $N$naturalness~\cite{Arkani-Hamed:2016rle}.  In all these cases, the DM can also be composed by a large number of species that could in principle contribute to lowering the gravitational cutoff of the theory.  The framework presented here fits nicely within this spectrum of ideas.  Our models are radical in their simplicity -- the hierarchy problem and the nature of DM could be deeply linked, with observable consequences that could show up in future collider and indirect detection experiments.

\section*{Acknowledgments}
\vspace{0.5pt}
We would like to thank A.\,\,Berlin, S.\,A.\,R.\,\,Ellis, M.\,\,Lisanti, J.\,\,Maldacena, G.\,\,Marques-Tavares, R.\,\,Rattazzi, and L.\,\,Vecchi for useful discussions.   TC is supported by the U.S.~Department of Energy, under grant number DE-SC0011640.  RTD is supported by the U.S.~Department of Energy under grant number DE-AC02-76SF00515.  ML acknowledges support from the Institute for Advanced Study.
\end{spacing}

\clearpage
\bibliography{LargeN_DM}
\bibliographystyle{utphys}

\end{document}